%% file: author.tex
\documentclass{svproc}
\usepackage{url}

\usepackage{graphicx}
\usepackage{multicol}
\usepackage{footmisc}

\pdfoutput=1

\input{./code}

\input{./concepts}

\input{./units}

\begin{document}
\mainmatter              % start of a contribution
\title{H-He Shell Interactions and Nucleosynthesis in Massive Population III Stars}
\titlerunning{Pop III H-He Shell Interactions}  % abbreviated title (for running head)
\author{Ondrea Clarkson\inst{1,4,5} \and Falk Herwig\inst{1,4,5}
Robert Andrassy\inst{1,4} \and Paul Woodward\inst{2,4} \and Marco Pignatari\inst{3,4,5} \and Huaqing Mao\inst{2,4}}
\authorrunning{Ondrea Clarkson et al.} % abbreviated author list (for running head)
%Craig Chambers, Kim B. Bruce, and Elisa Bertino}
%
\institute{Department of Physics \& Astronomy, University of Victoria,
 P.O. Box 3055 Victoria, B.C., V8W 3P6, Canada,\\
\email{oclark01@uvic.ca},\\ 
\and
LCSE and Department of Astronomy, University of Minnesota, Minneapolis, MN 55455, USA
\and
E.A. Milne Centre for Astrophysics, University of Hull, HU6 7RX, United Kingdom
\and
Joint Institute for Nuclear Astrophysics, Center for the Evolution of the Elements, Michigan State
State University, 640 South Shaw Lane, East Lansing, MI 48824, USA
\and
NuGrid collaboration, http://www.nugridstars.org}

\maketitle              % typeset the title of the contribution

\begin{abstract}
We report on our ongoing investigation into the nucleosynthetic and hydrodynamic nature of mixing at the interface between the H- and He-convection zones in massive Pop III stars. Studying recent a grid of 26 1D stellar evolution simulations with different mixing assumptions, we find that H-He interactions occur in 23/26 cases.  We demonstrate the nucleosynthesis expected in a H-He interaction in an 80M$_\odot$. Finally, we describe our progress in simulating a Pop III double convection zone in the \ppm hydrodynamics code.
% We would like to encourage you to list your keywords within
% the abstract section using the \keywords{...} command.
\end{abstract}
Pop III stars are thought to have produced and released the first elements heavier than those created in the Big Bang \cite{Nomoto13}. The most metal-poor stars we observe today may be the most direct descendants of Pop III stars and are a powerful diagnostic in our study of early cosmic chemical evolution \cite{Frebel15}. 

Interactions between H and He-convection layers have been seen in 1D stellar evolution simulations of massive Pop III stars \cite{2012ApJS..199...38L,heger:10,2001A&A...371..152M} but until recently, have not been investigated in detail. 

Similar convective-reactive events occur in additional environments, such as He-shell flashes in low-Z low-mass stars \cite{2012ASPC..458...45S,2004ApJ...602..377I,2014nic..confE.145D}, post-AGB stars \cite{2011ApJ...727...89H}, rapidly-accreting white dwarfs \cite{2017ApJ...834L..10D}, and low-Z Super-AGB stars \cite{2016MNRAS.455.3848J}. In these cases---just as Pop III stars--- likely leading to the \ipr \, with neutron densities $\approx \ee{13-15}\  \mathrm{cm^{-3}}$, first discussed by Cowan $\&$ Rose \cite{1977ApJ...212..149C}.

We have explored the possibility that the abundance patterns of the CEMP-no stars SMSS J031300, HE\,1327-2326 and HE\, 0107-5240, among the most iron-poor stars known, could be explained by highly energetic, convective H-He shell interactions in massive Pop-III stars \cite{2018MNRAS.474L..37C} without a strong odd-even effect. Based on a 45 $\Msun$ stellar model that undergoes a H/He convective-reactive event during C-core burning, we ran single-zone calculations to ascertain the nucleosynthesis which may result from such an event. For these simulations we found neutron densities of $\approx 6 \ee{13}\  \mathrm{cm^{-3}}$,  leading to striking similarities with the abundance patterns existing in some of the most metal-poor stars, particularly in abundance ratio trends seen from Na-Si and Ti-Mn.   

We have run a grid of 26 models using the \mesa\ stellar evolution code \cite{paxton:15}  over a mass range of $15 - 140\Msun$. For each initial mass, we use 5 different sets of mixing assumptions in order to explore the dependence of H-He interactions on marcophysical modelling choices. 
%Briefly, the choices are for each mass are as follows: a set of models with the Ledoux criterion for convection and semiconvetion with $\alpha = 0.05$ and no overshooting, and a set with overshooting values of $f_{ov} = 0.001$, $f_{ov} = 0.01$ and sets using the Schwarzchild criterion for each of the aforementioned overshooting values. 
Our findings indicate that there are three distinct modes for the interaction: firstly, as in Clarkson et al. \cite{2018MNRAS.474L..37C} a convective H and He-shell interaction. Secondly, a convective H-shell mixing into a convective He-core and thirdly, a convective H-shell mixing down into a radiative He-shell. 

\begin{figure}
  \includegraphics[width=0.8\textwidth]{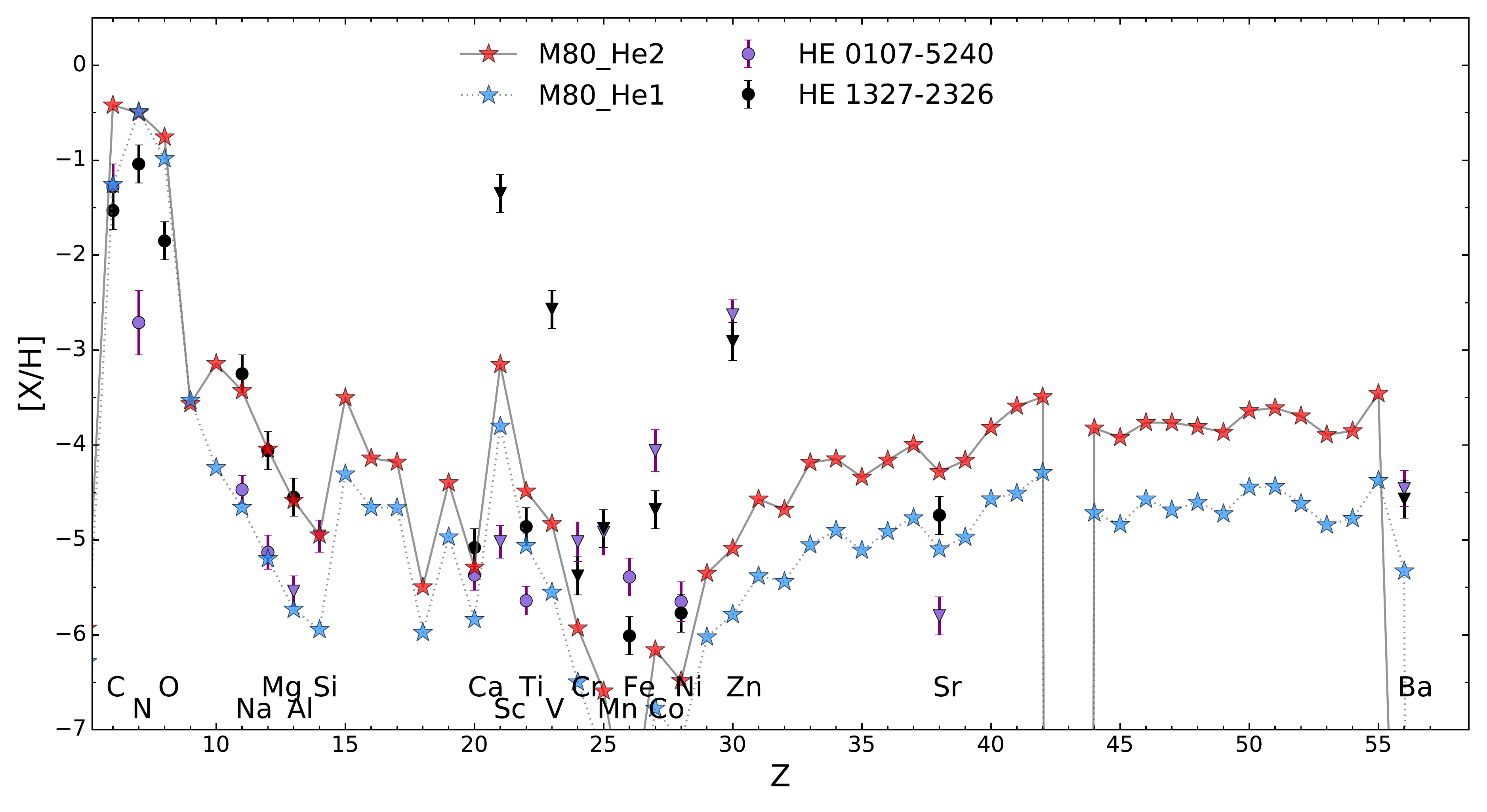}
 \caption{Abundances of CEMP-no stars HE\,0107-5240 and HE\,2317-2326 in purple and black are shown with single zone calculations (red and blue stars) based on an 80$\Msun$ stellar evolution simulation.}
 \label{ppn}
\end{figure}

Although difficult to constrain from 1D simulations, we have preformed additional single zone calculations more, with the aim to further explore the abundances of HE 0107-5240 and HE 2317-2326. We ran simulations with T = 2.5$\ee{8} \, \unitstyle{K}$ and $\rho$ = 1.9$\ee{2} \, \unitstyle{g \, cm^{-3}}$ from the He-shell of an 80$\Msun$ model from our grid of Pop III models with 1$\%$ H added, by mass. Preliminary results are shown in Fig. \ref{ppn} . We find that in order to simultaneously reproduce both light and trans-Fe elements in these stars the total neutron exposure must be a factor of ~4 smaller than we previously reported.

%Calcium has been reported to be made in massive Pop III stars via Hot CNO breakout reactions \cite{keller:14,2014ApJ...794...40T}, to explain the $\alpha$ elements in the most iron-poor stars without invoking explosive Si-burning.
%The three models that do not experience any H-He interaction, do not reproduce the Ca in SMSS J031300 by a factor of $\sim$10.This is because the H-burning temperatures are not high enough for any appreciable amount of time and the $^{19}\mathrm{F(p},\gamma)^{20}\mathrm{Ne}$ reaction  is a factor of $\sim10^4$ lower at these temperatures than the competing, $^{19}\mathrm{F(p},\alpha)^{16}\mathrm{O}$ rate. It remains unclear why this tension exists between the reports of different investigators and further investigation is necessary.

We have begun using the explicit PPMstar code \cite{2014ApJ...792L...3H} to investigate these events. Our initial suite of simulations contain the He-shell flash convection zone and the bottom of the H-burning convection zone, separated by a radiative zone of 25,000 km (Fig 2). Initially we are driving these convection zones by a constant volume heating at the bottom of the He convection zone and a corresponding cooling at the top of the H-burning convection zone. In order to realistically model this Pop III stellar environment, several code modifications were made. In future simulations, we will be including a simplified network to model the nuclear feedback expected from the mixing of H and He-burning material. The aim of these simulations is to answer the hypothesis \cite{2018MNRAS.474L..37C} that such an event may led to a GOSH-like instability and could potentially eject material from the star. We hope to determine whether such an event would occur, and if so, how would it unfold in a full 4$\pi$-3D environment in terms of possible asymmetries in the entrainment and how would the model respond to the nuclear feedback? 

\begin{figure}
  \includegraphics[width=1\textwidth]{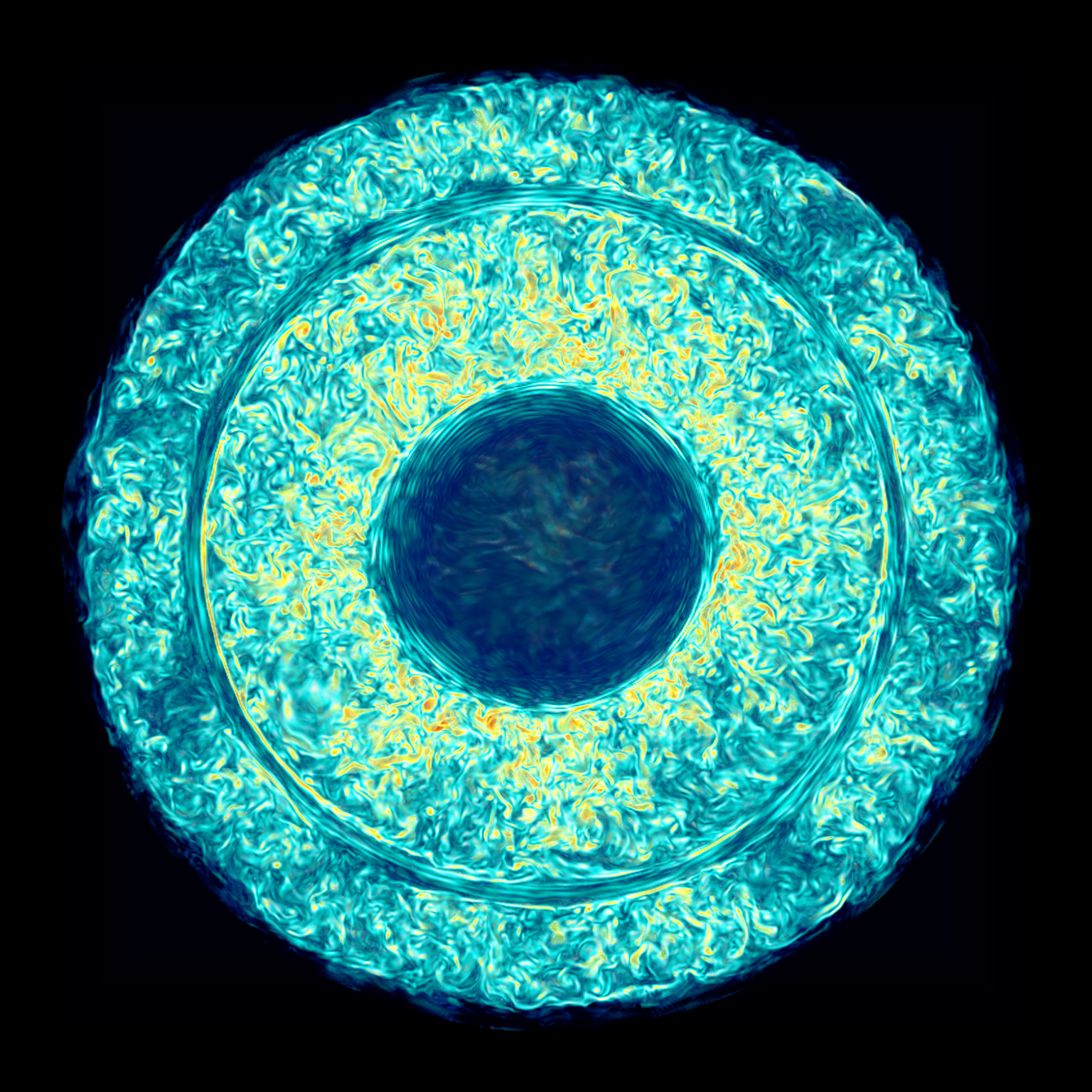}
 \caption{3D simulation of convective H and He-burning shells on a 768$^3$ grid. Only the first 50,000 km of the H-shell is simulated in order to adequately resolve the stable layer. Colours show vorticity.} 
 \label{ppm}
\end{figure}

The first message stars mayhave experienced violent convective-reactive interactions at the interface of the H- and He-burning regions. Three dimensional simulations with nuclear feedback are now being constructed to investigate this stellar and nuclear astrophysics  environment. A light-element i process could be triggered, and may result in abundance patterns observed in CEMP-no stars without strong odd-even effect.

\bibliographystyle{unsrt}
\bibliography{nic}

%\begin{thebibliography}{5}

%\end{thebibliography}{5}


\end{document}

%% file: code.tex
\newcommand{\code}[1]{\texttt{#1}}

\newcommand{\mesa}{\code{MESA}}

\newcommand{\ppm}{\code{PPMStar} }

%% file: concepts.tex
\newcommand{\ipr}{\mbox{$i$-process}}

%% file: units.tex
\newcommand{\unitstyle}[1]{\ensuremath{\mathrm{#1}}}

\newcommand{\ee}[1]{\ensuremath{\times 10^{#1}}}

\newcommand{\Msun}{\ensuremath{\unitstyle{M}_\odot}}